\documentclass[twocolumn, twoside, 10pt, a4paper]{IEEEtran}
 \usepackage{nopageno}
\usepackage{graphicx,color,epsf,bm}
\usepackage{amsmath,amsfonts,amssymb,amscd,bm}
\usepackage{flushend}
\usepackage{cuted}

\begin{document}
\title{\fontsize{17}{20}\bf
Analysis in $k$-space of Magnetization Dynamics Driven by Strong Terahertz Fields}%
\author{
\IEEEauthorblockN{V. Scalera\textsuperscript{1}, M. Hudl\textsuperscript{2}, K. Neeraj\textsuperscript{2}, S. Perna\textsuperscript{1}, M. d'Aquino\textsuperscript{3}, S. Bonetti\textsuperscript{2}, 
C. Serpico\textsuperscript{1}}
		\\
		\vspace{1ex}
{\textsuperscript{1}\small Department of Electrical Engineering and ICT, University of Naples Federico II, 80125 Naples, Italy}  \\
{\textsuperscript{2}\small Department of Physics, Stockholm University, 106 91 Stockholm, Sweden}  \\
{\textsuperscript{3}\small Department of Engineering, University of Naples "Parthenope", 80143 Naples, Italy} 
}
\maketitle

\begin{abstract}
Demagnetization in a thin film due to a terahertz pulse of magnetic field is investigated. Linearized LLG equation in the Fourier space to describe the magnetization dynamics is derived, and spin waves time evolution is studied. 
Finally, the demagnetization due to spin waves dynamics and recent experimental observations on similar magnetic system are compared. As a result of it, the marginal role of spin waves dynamics in loss of magnetization is established. 
\end{abstract}

\section{Introduction}
The mode of operation of magnetic storage technologies strongly relies on the control of fast
magnetization reorientation in a small region of a ferromagnetic media. Thus, in order to increase the efficiency
and the speed of these technologies, it is crucial to develop techniques to obtain increasingly faster magnetic
dynamics. In this respect, large research efforts are currently carried out to achieve fast magnetic reorientation
dynamics by using intense electromagnetic pulses \cite{Walowski}. In the early pioneering work in this area, femtosecond
optical pulses were used to induce magnetization dynamics indirectly via electronic excitation \cite{Beaurepaire, Kirilyuk}. More
recently, it has been shown that intense terahertz (THz) pulses can be used to achieve ultrafast magnetization dynamics by direct Zeeman coupling of magnetization with the magnetic field component of the
pulse \cite{Hudl}. It turns out that this technique enables to approach the fastest possible magnetization reversal \cite{Gamble}.  A surprising result of these experiments is the reduction of the magnetization module \cite{Hudl}, even when the THz pulses have energies too small to heat the medium. Such demagnetization process has been explained in terms of ultrafast scattering of spin polarized currents \cite{Bonetti}.\\
In this work, the role of spin waves dynamics in the demagnetization process is investigated.
In particular, the occurrence of inhomogeneities in the magnetization pattern due to spin waves excitation for a thin film excited by THz pulses, similar to the one considered in ref. \cite{Hudl}, is evaluated. The linear regime is considered and magnetization dynamics in terms of plane waves \cite{Suhl, Bertotti, Stancil} in the Fourier transform domain ($k$-space) is described. The dispersion relations are derived and the demagnetization effect due to the spin waves excitation is numerically computed and compared with experimental results. It is found that the role of spin waves induced inhomogeneities is several order of magnitude smaller than that measured in ref. \cite{Hudl}, highlighting the importance of spin-transport phenomena in ultrafast magnetization dynamics.

\section{Magnetization Dynamics}
\begin{figure}[h]
\includegraphics[width=.5\textwidth]{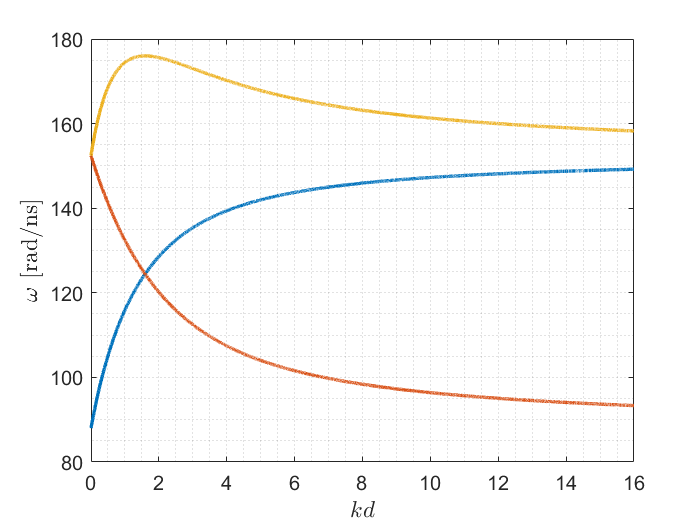}
\caption{\label{fig: magnetostatic waves} Dispersion relations of magnetostatic waves ($\ell_{EX}=0$). The saturation magnetization is $\mu_0M_S=1$ T and the applied field at the equilibrium is $\mu_0H_0=0.5$ T, i.e. the applied field is $\mu_0H_a=0.5$ T for the magnetization in plane and $\mu_0H_a=1.5$ T for the magnetization out of plane. The blue curve is represents \eqref{eq: forward}, the red and yellow curves represent \eqref{eq: backward} for $\phi_k=0$ and $\phi_k=\pi/2$ respectively.}
\end{figure}

The system considered is a thin film where the magnetization dynamics is assumed to be described by the LLG equation, expressed by the following equation:
\begin{equation}\label{eq: iLLG equation}
\frac{\partial\mathbf{M}}{\partial t}=\gamma\mathbf{M}\times\left(\mathbf{H}_\text{eff}-\frac{\alpha}{\gamma M_S}\frac{\partial\mathbf{M}}{\partial t}\right)\ ,
\end{equation}
where $\gamma$ is the gyromagnetic ratio, $\alpha$ is the damping constant, $\mathbf{M}(\mathbf{r},t)$ is the magnetization, $M_S$ is the saturation magnetization, $\mathbf{H}_\text{eff}(\mathbf{r},t)$ is the effective magnetic field.\\
The effective magnetic field is given by the sum of several contributions according to the following equation:
\begin{equation}\label{eq: effective field}
\mathbf{H}_\text{eff}=\mathbf{H_a}(\mathbf{r},t) +\mathbf{H}_M[\mathbf{M}]+\ell^2_{EX}\nabla^2\mathbf{M}\ ,
\end{equation}
where $\mathbf{H}_a$ is the applied field, $\ell_{EX}$ is the exchange length, and the demagnetizing field $\mathbf{H}_M$ is given by
\begin{equation}\label{eq: demagnetizing field int}
\mathbf{H}_M(\mathbf{r},t)=-\frac{1}{4\pi}\nabla\nabla\cdot\int_{\mathbb{R}^3}\frac{\mathbf{M}(\mathbf{s})}{|\mathbf{r}-\mathbf{s}|}\ \mathrm{d}^3{\mathbf{s}}\ .
\end{equation}
Let the applied magnetic field be
\begin{equation}
\mathbf{H}_a(\mathbf{r},t)=\mathbf{H}_{DC}+\mathbf{H}_\text{THz}(\mathbf{r},t)\ ,
\end{equation}
where $\mathbf{H}_{DC}$ is constant in time and space, and $\mathbf{H}_\text{THz}(\mathbf{r},t)$ is a pulse with sub-picosecond time duration.
In the following it is considered the case where the film thickness  $d$ satisfies the relation $d\sim\ell_{EX}$. This produces that the magnetization does not change significantly along the film thickness and then makes reasonable the following assumption:
\begin{equation}
\mathbf{M}(\mathbf{r},t) = \mathbf{M}(x,y,t) = \mathbf{M}(\bm\rho,t)\ ,
\end{equation}
where $\bm\rho\in\mathbb{R}^2$ is used to denote points of the $xy$-plane.\\
In this case only the average value of magnetic field (over the film thickness)  affects  the magnetization dynamics, namely
\begin{equation}
\left\langle\mathbf{H}_\text{eff}\right\rangle\,(\bm\rho,t) = \frac{1}{d}\int_{-d/2}^{+d/2}\mathbf{H}_{\text{eff}}(x,y,z,t)\ \mathrm{d}z\ .
\end{equation}
Similarly, we define $\langle\mathbf{H}_\text{THz}\rangle$ and $\langle\mathbf{H}_M\rangle$ as the average over the film thickness of $\mathbf{H}_\text{THz}$ and $\mathbf{H}_M$ respectively.\\ 
At this point, it is useful to express the magnetization vector field through its Fourier transform: 
\begin{equation}\label{eq: spin waves}
\mathbf{M}(\bm\rho,t)=\frac{1}{4\pi^2}\int_{\mathbb{R}^2} \mathcal{M}(\mathbf{k},t)\exp\left(i\mathbf{k}\cdot\bm\rho\right)\ \mathrm{d}^2\mathbf{k}\ ,
\end{equation}
which expresses the fact that the magnetization is seen as a continuous superposition of spin waves.
When we apply the demagnetizing field and the exchange field operator to the single spin wave, we obtain an other spin wave with the same wave vector of the magnetization. In fact we have:
\begin{equation}\label{eq: demagnetizing field}
\left\langle\mathbf{H}_M\right\rangle\left[\mathcal{M}(\mathbf{k},t)e^{i\bm\rho\cdot\mathbf{k}}\right]=-D(k)\,\mathcal{M}(\mathbf{k},t)\ e^{i\bm\rho\cdot\mathbf{k}}
\ ,
\end{equation}
and
\begin{equation}\label{eq: exchange field}
\ell^2_{EX}\nabla^2\left(\mathcal{M}(\mathbf{k})e^{i\bm \rho\cdot\mathbf{k},t}\right)=-k^2\ell_{EX}^2\left(\mathcal{M}(\mathbf{k},t)e^{i\bm\rho\cdot\mathbf{k}}\right)
\, ,
\end{equation}
where $k=||\mathbf{k}||$ and $D(k)$ is a symmetric $3\times3$ matrix (see appendix \ref{app: demagnetizing factors}).\\
As we will see in the following, equations \eqref{eq: demagnetizing field} and \eqref{eq: exchange field} imply the well known fact that the linear dynamics of spin waves with different wave vectors are independent.
When $\mathbf{H}_\text{THz}(t)=\mathbf{0}$, the equilibrium condition of \eqref{eq: iLLG equation} is satisfied by uniform magnetization and is expressed by the following Brown's equation
\begin{equation}\label{eq: equilibrium}
\mathbf{M}_0\times\langle\mathbf{H}_{\text{eff},0}\rangle=\bm 0\ .
\end{equation}
which in scalar form reads as
\begin{equation}
H_{DC}\sin(\theta_H-\theta_M)-M_S\cos\theta_M\sin\theta_M=0
\ ,
\end{equation}
where $\theta_M$ and $\theta_H$ are the angles between the unit vector $\bm e_z$ and the vectors $\mathbf{M}_0$ and $\mathbf{H}_{DC}$ respectively.\\
Introducing a Cartesian coordinate system $\lbrace\bm e_m,\bm e_\theta, \bm e_\varphi\rbrace$, given by
\begin{equation}\label{eq: reference frame}
\bm e_m = \frac{\mathbf{M}_0}{M_S}\ ,\quad
\bm e_{\varphi} = \frac{\bm e_z\times\bm e_m}{|\bm e_z\times\bm e_m|}\ ,\quad
\bm e_{\theta}=\frac{\bm e_\varphi\times\bm e_m}{|\bm e_\varphi\times\bm e_m|}\ ,
\end{equation}
the magnetization is written as
\begin{equation}
\mathbf{M}(\bm\rho,t)=M_S\ \bm e_m+M_\theta(\bm\rho,t)\ \bm e_\theta+M_\varphi(\bm\rho,t)\ \bm e_\varphi,
\end{equation}
where $M_\theta(\bm \rho,t)$ and $M_\varphi(\bm\rho,t)$ are the first order variations.
By linearizing  \eqref{eq: iLLG equation} and projecting it on the plane orthogonal to $\mathbf{M}_0$, it eventually yields
\begin{equation}\label{eq: linear iLLG t}
\frac{\partial M_\theta}{\partial t}-\alpha\frac{\partial M_\varphi}{\partial t}=\gamma H_0M_\varphi-\gamma M_S H_{\varphi}\ ,
\end{equation}
\begin{equation}\label{eq: linear iLLG p}
-\alpha\frac{\partial M_\theta}{\partial t}-\frac{\partial M_\varphi}{\partial t}=\gamma H_0M_\theta-\gamma M_S H_{\theta}\ ,\ \ 
\end{equation}
where 
\begin{equation*}
H_\varphi=\left\langle\mathbf{H}_\text{eff}\right\rangle\cdot\bm e_\varphi\ ,\ \ H_\theta=\left\langle\mathbf{H}_\text{eff}\right\rangle\cdot\bm e_\theta\ ,\ \ H_0 = ||\langle\mathbf{H}_{\text{eff},0}\rangle||\ .
\end{equation*}
In this framework the field $\langle\mathbf{H}_{THz}\rangle(\bm\rho,t)$  is treated as a first order perturbation.\\ 
Replacing expression \eqref{eq: effective field} of the effective field (along with \eqref{eq: exchange field} and \eqref{eq: demagnetizing field} ) into \eqref{eq: linear iLLG t} and \eqref{eq: linear iLLG p} and taking the Fourier transform in space, we arrive to the following system of equations:
\begin{equation}\label{eq: main model}
\frac{\partial}{\partial t}
\left[ 
\begin{array}{c}
\mathcal{M}_\theta\\\mathcal{M}_\varphi
\end{array}
\right]=A(\mathbf{k})
\left[
\begin{array}{c}
\mathcal{M}_\theta\\\mathcal{M}_\varphi
\end{array}
\right]-\gamma M_S
\left[
\begin{array}{c}
\mathcal{H}_\theta\\\mathcal{H}_\varphi
\end{array}
\right]
\ ,
\end{equation}
where  $\mathcal{H}_\theta$ and $\mathcal{H}_\varphi$ are the components along $\bm e_\theta$ and $\bm e_\varphi$ of the Fourier transform in space of the applied THz field, the dynamical matrix is 
\begin{equation}\label{eq: dynamical matrix}
A= 
\left[
\begin{array}{cc}
1&-\alpha\\-\alpha&-1
\end{array}
\right]^{-1}
\left[
\begin{array}{cc}
\gamma M_S D_{\theta\varphi}(\mathbf{k})&\gamma \hat{H}_\theta(\mathbf{k})\\
\gamma \hat{H}_\varphi(\mathbf{k})&\gamma M_S D_{\varphi\theta}(\mathbf{k})
\end{array}
\right]\ ,
\end{equation}
and
\begin{equation}\label{eq: H hat}
\hat{H}_i(\mathbf{k})=H_0+M_SD_{ii}(\mathbf{k})+M_Sk^2\ell_{EX}^2\ .
\end{equation}
For every $\mathbf{k}$, equation \eqref{eq: main model} describes a $2\times2$ linear system without interactions between spin waves with different $\mathbf{k}$. Dynamics can be simulated separately for every wave vector, whereas the magnetization in real space is obtained through \eqref{eq: spin waves}.\\
At this point, it is useful to derive the dispersion relations for spin waves dynamics, which will be of help in explaining the numerical results of the next section.
Dispersion relations are obtained by imposing $\det(A-i\omega I)=0$  with $\alpha=0$.\\
Two cases are considered: first, when the exchange field in negligible compared with the demagnetizing field and second, when the demagnetizing field is negligible compared to the exchange field.\\
Let us start with the first case. This situation occurs when the demagnetizing coefficients $D_{ij}(\mathbf{k})$ are much greater than $\ell_{EX}^2k^2$, since $|D_{ij}|\leq 1$ and then the condition required is $\ell_{EX}^2k^2\ll 1$.\\
This type of waves are called magnetostatic waves.\\
Let us focus on the case with the magnetization out of plane ($\theta_M=0$). This occurs when the applied field is out of plane and $H_a>M_S$. By replacing $H_0=H_a-M_S$ and the demagnetizing factors with their expressions in \eqref{eq: D expressions}, it eventually yields
\begin{equation}\label{eq: forward}
\omega=-\gamma\sqrt{(H_a-M_S)(H_a-M_SS_k)}\ ,
\end{equation}
where $S_k$ is defined in \eqref{eq: S(k)}. The dispersion relation \eqref{eq: forward} has a positive group velocity and it is called forward magnetostatic wave \cite{Stancil}.\\
Next we consider the magnetization is in plane ($\theta_M=\pi/2$), which occurs when the applied field is in plane. Replacing $H_0 = H_a$ and the expressions of the demagnetizing factors from \eqref{eq: D expressions} yields
\begin{equation}\label{eq: backward}
\omega=-\gamma\sqrt{(H_a+M_SS_k)(H_a+M_S(1-S_k)\sin^2\phi_k)}\ ,
\end{equation}
where $\phi_k$ is the angle between $\mathbf{M}_0$ and $\mathbf{k}$. When $\phi_k=0$, \eqref{eq: backward} has a negative group velocity and it called backward magnetostatic wave \cite{Stancil}.\\
The plot of \eqref{eq: forward} and \eqref{eq: backward} are in Figure \ref{fig: magnetostatic waves}.\\
Consider now the case $\ell_{EX}^2k^2\gg 1$, i.e. the magnetostatic field is negligible. In this case the dispersion relation does not depend on the orientation $\mathbf{k}$. We have
\begin{equation}
\omega = -\gamma(H_0+M_S\ell_{EX}^2k^2)\ .
\end{equation}
 For spatially uniform, or nearly uniform, terahertz pulse distribution $(k d \ll 1)$ the resonance frequencies for spin waves excitation are in the order of $\sim \,10\,GHz$. Therefore, it is expected that the nonlinear spin waves dynamics regime is not reached, even for high power magnetic field pulse used for the experimental investigations of ref. \cite{Hudl}. 
\section{Simulation}

\begin{figure}[h]
\includegraphics[width=.5\textwidth]{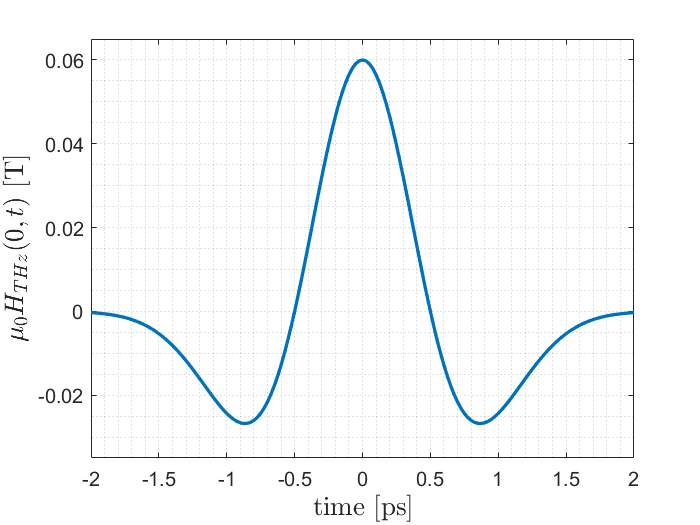}
\caption{\label{fig: HTHZ} Time profile of the applied terahertz pulse.}
\end{figure}

\begin{figure}[h]
\includegraphics[width=.5\textwidth]{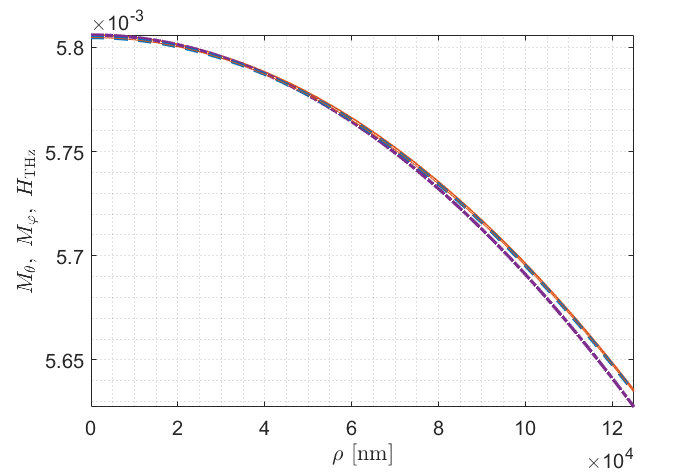}
\caption{\label{fig: space gauss} Space dependence of the applied terahertz field $H_\text{THz}$ and the perturbation of the magnetization $M_\theta$ and $M_\varphi$. The red solid line is $M_\theta$, the dashed blue line is $M_\varphi$ and the dashed purple line is $H_\text{THz}$. The plot are rescaled so that the in $\rho=0$ the three curves have the same value.}
\end{figure}

\begin{figure*}
\includegraphics[width=.5\textwidth]{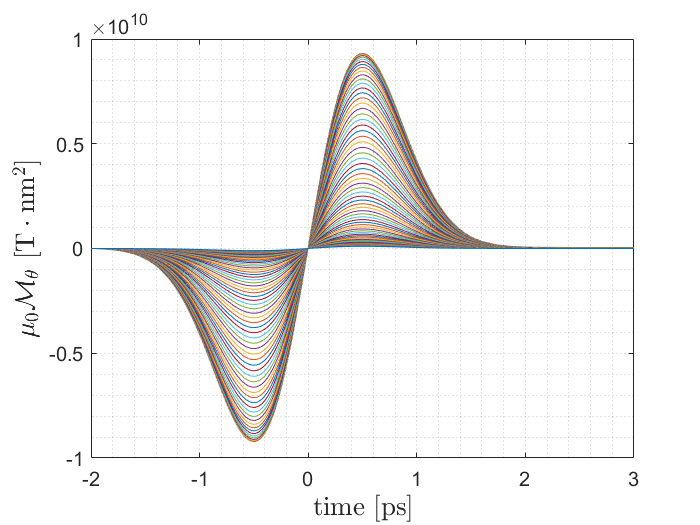}
\includegraphics[width=.5\textwidth]{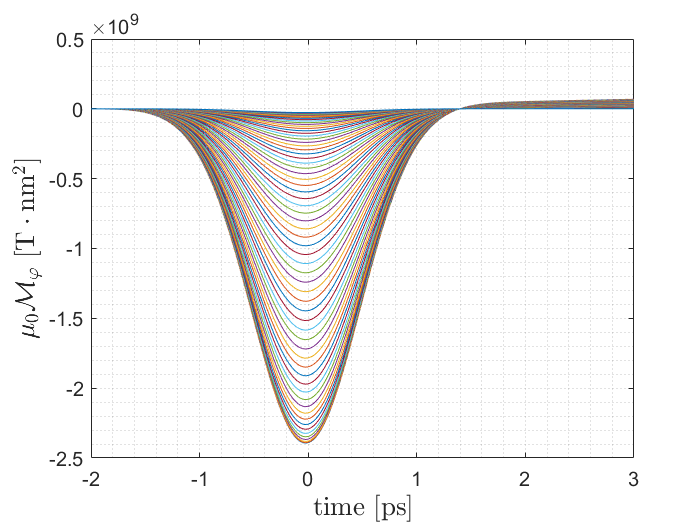}
\caption{\label{fig: simulation} Amplitude of the spin waves of the system described in the simulation section. Lager curves correspond to spin waves with smaller wave number, the orientation of $\mathbf{k}$ has almost no impact on the dynamic.}
\end{figure*}

We consider a system similar to the one used in \cite{Hudl}: The specimen is a thin film with thickness $d=5$ nm, the material parameters are $\mu _0M_S=1.84$ T, $\gamma=-176\ \mathrm{rad}/(\mathrm{T\cdot ns})$, $\alpha=-0.007$ and $\ell_{EX}=4.4$ nm, and the applied magnetic field out of plane component $\mu_0H_\perp=0.448$ T whereas the in plane component is $\mu_0H_\parallel=0.056$ T. 
The terahertz field is applied in plane (orthogonally to the constant field) and the intensity is
\begin{equation}
H_{THz}(\bm\rho,t) = \frac{H_\text{Max}}{\sigma_t^2}(t^2-\sigma_t^2)\exp\left(-\frac{t^2}{2\sigma_t^2}-\frac{\rho^2}{2\sigma_x^2}\right)\ ,
\end{equation}
where $\mu_0H_\text{Max}=0.06$ T, $\sigma_x = 500\ \mu\text{m}$ and $\sigma_t = 0.5$ ps. The time evolution of $H_{THz}$ is displayed in the left panel of Figure \ref{fig: HTHZ}.\\
The system \eqref{eq: main model} is simulated within a range of wave number up to $10^{-5}\ \text{nm}^{-1}$,  higher wave numbers are not excited by the applied field. The time evolution of $\mathcal{M}(\mathbf{k},t)$ is shown Figure \ref{fig: simulation} for several values of $\mathbf{k}$.\\
In the range of wave vector excited the dynamics does not change appreciably, hence the space profile of either $\mathcal{M}_\theta$ and $\mathcal{M}_\varphi$ are almost the same as $H_\text{THz}(\bm\rho,0)$ rescaled. The space dependence of $\mathcal{M}_\theta$ and $\mathcal{M}_\varphi$ are in Figure \ref{fig: space gauss}.

\begin{figure}
\includegraphics[width=.5\textwidth]{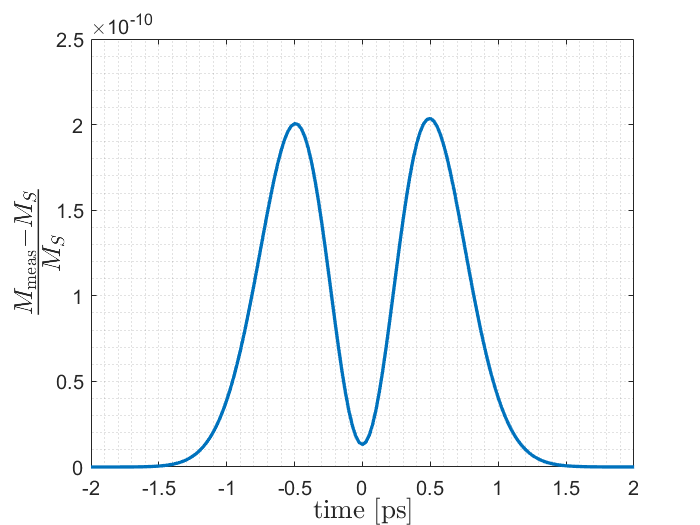}
\caption{\label{fig: demag} Relative reduction of the measured magnetization.}
\end{figure}

In order to evaluate the demagnetization observed the magnetization is normalized, since the linearized model does not preserve the magnetization modulus. We have
\begin{equation}\label{eq: demag}
\mathbf{M} = M_S\frac{\mathbf{M}_0+M_\theta\bm e_\theta+M_\varphi\bm e_\varphi}{||\mathbf{M}_0+M_\theta\bm e_\theta+M_\varphi\bm e_\varphi||}=M_S\frac{\mathbf{M}_0+\delta\mathbf{M}}{||\mathbf{M}_0+\delta\mathbf{M}||}\ ,
\end{equation}
where $\delta\mathbf{M}=M_\theta\bm e_\theta+M_\varphi\bm e_\varphi$.\\
The measured magnetization is given by 
\begin{equation}
M_\text{meas} = \frac{1}{|\Omega_M|}\left\lVert\ \int_{\Omega_M}\mathbf{M}\ \mathrm{d}S\ \right\rVert
\end{equation}
where $\Omega_M$ is the area hit by the probe, i.e. a circular area with radius $125\ \mu$m, and $|\Omega_M|$ is the measure of the area.\\
Developing the Taylor series up to the second order in $\delta\mathbf{M}/M_S$ and using $\mathbf{M}_0\perp\delta\mathbf{M}$,  the integral is approximated by
\begin{equation}\label{eq: Taylor}
\int_{\Omega_M}\mathbf{M}\ \mathrm{d}S\approx\int_{\Omega_M}\left(\mathbf{M}_0-\frac{\mathbf{M}_0}{2}\left\lVert\frac{\delta\mathbf{M}}{M_S}\right\rVert^2+\delta\mathbf{M}\right)\ \mathrm{d}S\ .
\end{equation}
By replacing \eqref{eq: Taylor} in \eqref{eq: demag}, neglecting the terms of order greater than two in $\delta\mathbf{M}/M_S$, we eventually obtain the demagnetization
\begin{equation}
M_\text{meas} = \sqrt{M_S^2-I_2+I_1}\ ,
\end{equation}
where
\begin{equation}
I_1=\frac{1}{|\Omega_M|^2}\left(\int_{\Omega_M}\delta\mathbf{M}\ \mathrm{d}S\ \right)^2\ ,
\end{equation}
and
\begin{equation}
I_2 = \frac{1}{|\Omega_M|}\int_{\Omega_M}\delta\mathbf{M}^2\ \mathrm{d}S\ .
\end{equation}
It is noteworthy that $I_2\geq I_1$ because of Cauchy-Schwartz inequality, hence the measured magnetization can only be smaller than $M_S$.\\ 
The plot of the relative reduction of the measured magnetization is displayed in Figure \ref{fig: demag}.\\
Remarkably, the reduction of the observed magnetization does not grow over $2.5\cdot10^{-10}$, whereas the demagnetization observed in \cite{Hudl} is several order of magnitude higher (roughly $2\cdot 10^{-3}$).\\

\section{Conclusion}
In conclusion, the demagnetization effect induced in a thin film excited by a strong terahertz pulse is investigated. The magnetization dynamics is described in terms of linear spin waves dynamics governed by the linearized LLG projected into the Fourier's $k$-space.
In this micromagnetic framework, we tried to reproduce the experiment described in \cite{Hudl} and determine whether the spin waves play a role in the experimental observed demagnetization explained in terms of ultrafast scattering of spin polarized currents \cite{Bonetti}. It is found, that the reduction of the magnetization module computed in the simulation is several orders of magnitudes smaller than the demagnetization observed in the experiments. Moreover, despite the field intensity is high, the pulse is so short that the magnetization barely moves from equilibrium and spin waves do not grow enough to cause relevant nonlinear effects \cite{Suhl, Bertotti}.  Then, the magnetization reduction  due to non-uniformities is negligible and the role of ultrafast spin-transport phenomenon is prevalent.\\

\appendices
\section{Demagnetizing Coefficients Computation}
\label{app: demagnetizing factors}
\begin{figure}
\centering
\includegraphics[width=.5\textwidth]{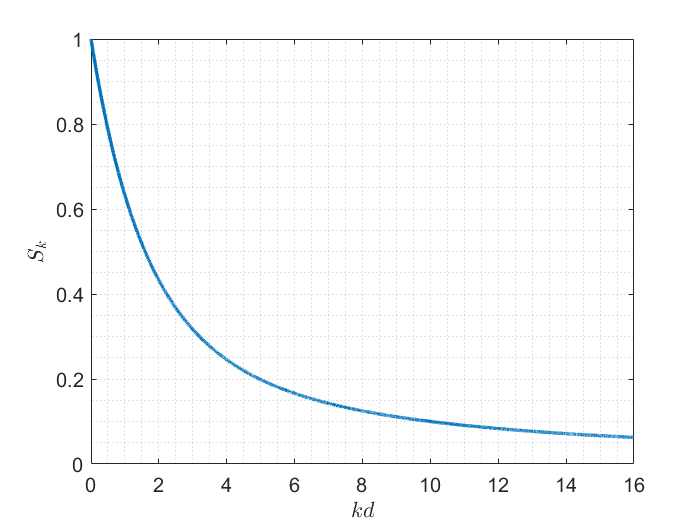}
\caption{\label{fig: Sk} $S_k$ function defined in \eqref{eq: S(k)}}
\end{figure}

This section derives the analytical expressions for the demagnetizing factor $D(\mathbf{k})$ defined in equation \eqref{eq: demagnetizing field}. Its elements express the linear relation between the averaged demagnetizing field and the magnetization
\begin{equation}
\left[
\begin{array}{c}
\langle H_{Mx}\rangle\\\langle H_{My}\rangle\\\langle H_{Mz}\rangle
\end{array}
\right]=
-\left[
\begin{array}{ccc}
D_{xx}&D_{xy}&D_{xz}\\D_{xy}&D_{yy}&D_{yz}\\D_{zx}&D_{zy}&D_{zz}
\end{array}
\right]
\left[
\begin{array}{c}
M_{x}\\M_{y}\\M_{z}
\end{array}
\right]
\, .
\end{equation}
The $D_{ij}(\mathbf{k})$  can be obtained from \eqref{eq: demagnetizing field int}, which is the general solution of the magnetostatic problem
\begin{equation}\label{eq: magnetostatic 1}
\nabla\times\mathbf{H}_M=0\ ,
\qquad
\nabla\cdot\mathbf{H}_M=-\nabla\cdot\mathbf{M}\ .
\end{equation}
Alternatively, the system \eqref{eq: magnetostatic 1} can be restated in terms of a scalar potential $\psi$, that is
\begin{equation}\label{eq: magnetostatic 2}
\mathbf{H}_M=-\nabla\psi\ ,\qquad \nabla^2\psi=\nabla\cdot\mathbf{M}\ .
\end{equation}
Let us focus on the demagnetizing field in the thin film with  magnetization given by
\begin{equation}\label{eq: spin wave thin film}
\mathbf{M}(\mathbf{r}) =
\begin{cases}
\mathbf{M}_0 \exp(i\mathbf{k}\cdot\bm\rho)\qquad\text{if}\quad|z|\leq d/2
\\
\ 0\qquad\qquad\qquad\quad\ \,\text{if}\quad|z|>d/2
\end{cases}.
\end{equation}
The particular solution of \eqref{eq: magnetostatic 2} is
\begin{equation}\label{eq: potential - particular solution}
\psi_p(\mathbf{r}) =
\begin{cases}
\dfrac{\mathbf{k}\cdot\mathbf{M}_0}{ik^2}\,\exp(i\mathbf{k}\cdot\bm\rho)\qquad\text{if}\quad|z|\leq d/2
\\
\ \quad0\qquad\qquad\qquad\qquad\ \text{if}\quad|z|>d/2
\end{cases}.
\end{equation}
The potential \eqref{eq: potential - particular solution} does not satisfy the interface conditions on the thin film surfaces, namely the continuity of the normal component of the magnetic density flux
\begin{equation}\label{eq: interface n}
(\mathbf{H}_M^m(\mathbf{r},t)+\mathbf{M}(\mathbf{r},t))\cdot\bm e_z=\mathbf{H}_M^a(\mathbf{r},t)\cdot\bm e_z\ ,
\end{equation}
and the continuity of the tangent component of the magnetic field
\begin{equation}\label{eq: interface t}
\mathbf{H}_M^m(\mathbf{r},t)\times\bm e_z=\mathbf{H}_M^a(\mathbf{r},t)\times\bm e_z\ ,
\end{equation}
where the superscript `\textit{m}' and `\textit{a}' denote the 
fields in the magnetic medium and in the air respectively, and equations \eqref{eq: interface n} and \eqref{eq: interface t} are imposed at $z=\pm d/2$.\\
The solution of \eqref{eq: magnetostatic 2} is obtained by summing to \eqref{eq: potential - particular solution} a linear combination of harmonic functions
\begin{equation}
\psi_{\pm}(\mathbf{r}) = \exp(i\mathbf{k}\cdot\bm\rho\pm kz)\ .
\end{equation}
The homogeneous solution is 
\begin{equation}
\psi_0(\mathbf{r}) =
\begin{cases}
C_1\psi_{-}\qquad\qquad\qquad\text{if}\ \ z\ >\ d/2
\\
C_2\psi_{-}+C_3\psi_{+}\qquad\ \text{if}\ |z|\leq\ d/2
\\
C_4\psi_{+}\qquad\qquad\qquad\text{if}\ \ z\ <-d/2
\end{cases}\ ,
\end{equation}
where $C_1$, $C_2$, $C_3$ and $C_4$ are constant to be determined.\\
By using \eqref{eq: interface n} and \eqref{eq: interface t}, we obtain a set of linear equations in $C_1$, $C_2$, $C_3$ and $C_4$, which yields
\begin{equation}
\begin{aligned}
C_1 = -\frac{\sinh(kd/2)}{k}\ [(\mathbf{M}_0\cdot\bm e_k)+(\mathbf{M}_0\cdot\bm e_z)]\ ,
\\
C_2 = \frac{\exp(-kd/2)}{2k}\ [(\mathbf{M}_0\cdot\bm e_k)+(\mathbf{M}_0\cdot\bm e_z)]\ ,
\\
C_3 = \frac{\exp(-kd/2)}{2k}\ [(\mathbf{M}_0\cdot\bm e_k)-(\mathbf{M}_0\cdot\bm e_z)]\ ,
\\
C_4 = -\frac{\sinh(kd/2)}{k}\ [(\mathbf{M}_0\cdot\bm e_k)-(\mathbf{M}_0\cdot\bm e_z)]\ ,
\end{aligned}
\end{equation}
where $\bm e_k=(1/k)\mathbf{k}$.\\
The derivation of the averaged demagnetizing field is straightforward. We eventually have
\begin{equation}
\langle\mathbf{H}_M\rangle = -[(\mathbf{M}_0\cdot\bm e_z)S_k+(\mathbf{M}_0\cdot\bm e_k)(1-S_k)]\exp(i\mathbf{k}\cdot\bm\rho)\ ,
\end{equation}
where 
\begin{equation}\label{eq: S(k)}
S_k=[1-\exp(-kd)]/kd\ .
\end{equation}
The function $S_k$ is shown in Figure \ref{fig: Sk}.
By using the rotated reference frame defined in \eqref{eq: reference frame}, we obtain the demagnetizing factors used in \eqref{eq: dynamical matrix} and \eqref{eq: H hat}. They are
\begin{equation}\label{eq: D expressions}
\begin{aligned}
&D_{\theta\theta} = (1-S_k)\cos^2\theta_M\cos^2\phi_k+S_k\sin^2\theta_M\ ,
\\
&D_{\theta\varphi}=D_{\varphi\theta}=\frac{1}{2}(1-S_k)\cos\theta_M\sin(2\phi_k)\ ,
\\
&D_{\varphi\varphi}=(1-S_k)\sin^2\phi_k\ ,
\end{aligned}
\end{equation}
where $\theta_M$ is the angle between $\bm e_z$ and $\bm e_m$, and  $\phi_k$ is angle between $\bm e_k$ and the projection of either $\bm e_\theta$ or $\bm e_m$ in the $xy$-plane.


\begin{thebibliography}{}

\bibitem{Walowski}
 J. Walowski and M. Münzenberg,, ``Perspective: Ultrafast magnetism and THz spintronics",
 \textit{Journal of Applied Physics}, 
 vol. 120, 140901 (2016).
 
\bibitem{Beaurepaire}
 E. Beaurepaire, J. Merle, A. Daunois, J. Bigot, ``Ultrafast Spin Dynamics in Ferromagnetic Nickel",
 \textit{Physical Review Letters}, 
 vol. 76, pp. 4250-4253 (1996).
 
 \bibitem{Kirilyuk}
 A. Kirilyuk, A. V. Kimel, T. Rasing, ``Ultrafast optical manipulation of magnetic order",
 \textit{Reviews of Modern Physics}, 
 vol. 82, pp. 2731-2784, (2016).

\bibitem{Hudl}
M. Hudl, M. d’Aquino, C. Serpico, M. Pancaldi, S.-H. Yang, M.G. Samant, S.S.P. Parkin, H.A. Durr, M.C. Hoffmann, S. Bonetti, ``Nonlinear magnetization dynamics driven by strong terahertz fields", \textit{Physical Review Letters}, vol. 123 , issue 19, pp 197-204 (2019)

\bibitem{Bonetti}
S. Bonetti, M. C. Hoffmann, M. J. Sher, Z. Chen, S. H. Yang, M. G. Samant, S. S. P. Parkin, and H. A. Durr, ``THz-Driven Ultrafast Spin-Lattice Scattering in Amorphous Metallic Ferromagnets", \textit{Physical Review Letters}, 
vol. 117, 087205, (2016).

  \bibitem{Gamble}
  S. J. Gamble, M. H. Burkhardt, A. Kashuba, R. Allenspach, S. S. P. Parkin, H. C. Siegmann, and J. Stöhr, ``Electric Field Induced Magnetic Anisotropy in a Ferromagnet",
 \textit{Physical Review Letters}, 
 vol.  102, 217201 (2009).

\bibitem{Suhl}
H. Suhl, ``The theory of ferromagnetic resonance at high signal powers",
\textit{Journal of Physics and Chemistry of Solids}, 
vol. 1, issue 4, pp. 209-227 (1957)

\bibitem{Bertotti}
G. Bertotti, I.D. Mayergoyz, C. Serpico, ``Spin-Wave Instabilities in Large-Scale Nonlinear Magnetization Dynamics", 
\textit{Physical Review Letters}, 
vol. 87, issue 21, pp. 217203 (2001)

\bibitem{Stancil}
D.D. Stancil, A Prabhakar, ``Spin Waves: Theory and Applications", Springer (2009)


\end{thebibliography}
\end{document}